\documentclass[aps,twocolumn,prl,superscriptaddress,letterpaper]{revtex4-1}
\usepackage{amssymb}
\usepackage{mathtools}
\usepackage[pdftex]{graphicx}
\usepackage[usenames, dvipsnames, svgnames, table]{xcolor}
\usepackage[colorlinks=true, citecolor=Blue,
            linkcolor=BrickRed, urlcolor=ForestGreen]{hyperref}
\usepackage{txfonts}
\usepackage{mathrsfs}
\usepackage{bm}
\usepackage{multirow}
\usepackage{color}
\usepackage{comment}
\usepackage{CJK}

\newcommand{\Com}[1]{{\color{red}{#1}\normalcolor}} 

\begin{document}
\begin{CJK*}{UTF8}{gbsn}

\title{Maxwell Demon and Einstein-Podolsky-Rosen Steering}

\author{Meng-Jun Hu (\CJKfamily{gbsn}胡孟军)}
\email{humj@baqis.ac.cn}
\affiliation{Beijing Academy of Quantum Information Sciences, Beijing 100193, China}

\author{Xiao-Min Hu (\CJKfamily{gbsn}胡晓敏)}
\affiliation{Laboratory of Quantum Information, University of Science and Technology of China, Hefei 230026, China }%
\affiliation{Synergetic Innovation Center of Quantum Information and Quantum Physics,
University of Science and Technology of China, Hefei 230026, China}

\author{Yong-Sheng Zhang (\CJKfamily{gbsn}张永生)}%
\email{yshzhang@ustc.edu.cn}
\affiliation{Laboratory of Quantum Information, University of Science and Technology of China, Hefei 230026, China }%
\affiliation{Synergetic Innovation Center of Quantum Information and Quantum Physics,
University of Science and Technology of China, Hefei 230026, China}
\affiliation{Hefei National Laboratory, University of Science and Technology of China, Hefei, 230088, China}

\date{\today}

\begin{abstract}
The study of Maxwell demon and quantum entanglement is important because of its foundational significance in physics and its potential applications in quantum information. Previous research on the Maxwell demon has primarily focused on thermodynamics, taking into account quantum correlations. Here we consider from another perspective and ask whether quantum non-locality correlations can be simulated by performing work. The Maxwell demon-assisted Einstein-Podolsky-Rosen (EPR) steering is thus proposed, which implies a new type of loophole. The application of Landauer's erasure principle suggests that the only way to close this loophole during a steering task is by continuously monitoring the heat fluctuation of the local environment by the participant. We construct a quantum circuit model of Maxwell demon-assisted EPR steering, which can be demonstrated by current programmable quantum processors, such as superconducting quantum computers. Based on this quantum circuit model, we obtain a quantitative formula describing the relationship between energy dissipation due to the work of the demon and quantum non-locality correlation.
The result is of great physical interest because it provides a new way to explore and understand the relationship between quantum non-locality, information, and thermodynamics.
\end{abstract}

\maketitle
\end{CJK*}

\section{Introduction}

Maxwell demon was first proposed by James Clark Maxwell in 1867 to demonstrate that the second law of thermodynamics is statistical rather than based on dynamical laws such as those of Newton \cite{demon}. The Maxwell demon paradox was completely resolved by Landauer in 1961 when he introduced the concept of {\it logical irreversibility} for the process of memory erasure \cite{Landuer}. Landauer's erasure principle states that information erasure is a logically irreversible process in which energy dissipation must be involved, thus causing an entropy increase in the environment. Due to its important role in revealing connections between thermodynamics and information theory, the Maxwell demon has now been widely investigated in quantum thermodynamics \cite{thermo} and quantum information theory \cite{information1, information2}. 

Quantum entanglement is a fundamental characteristic of quantum theory and plays an important role as a resource in quantum information tasks \cite{nonlocality1, nonlocality2, nonlocality3}. 
From an information-theoretic perspective, the information entropy of an entangled system is lower than the mixed ones. It is natural to consider the connection between quantum entanglement and Maxwell demon \cite{entropy}. The study of quantifying the amount of entanglement and quantumness of correlations by using Maxwell demon \cite{Hodek,zurek, vlat}, and how much work that can be extracted from a correlated pair \cite{work1, work2, work3} has attracted significant interest in the past two decades. Meanwhile, research on quantum Maxwell demon and heat engine is still in progress \cite{engine1, engine2, engine3, engine4, engine5}. 
Here, we take a different perspective and ask whether the quantum non-locality correlation can be simulated by the Maxwell demon. In the context of Einstein-Podolsky-Rosen (EPR) steering, we demonstrate the existence of a protocol in which Alice and the demon collaborate to cheat Bob. This form of Maxwell demon-assisted EPR steering implies that non-locality correlations, which refute the local hidden state model, can be simulated through work. It also means the existence of a new type of loophole in the EPR steering which we call {\it Maxwell demon loophole}. By applying Landauer's erasure principle, we demonstrate that the only method to close this loophole involves Bob continuously monitoring the temperature fluctuation of the local environment. We construct the quantum circuit model for this Maxwell demon-assisted EPR steering, which is feasible to be demonstrated with current quantum processors. Based on this quantum circuit model, we derive the explicit formula that describes the relationship between quantum non-locality demonstrated in EPR steering and the work done by the demon.

This work is organized as follows: We first introduce the Maxwell demon-assisted EPR steering and discuss the physical limitations of the Maxwell demon that could potentially close this new type of loophole. We then construct the quantum circuit model of Maxwell demon-assisted EPR steering and derive the formula describing the relation between the work done by the demon and quantum non-locality based on this model. Further discussion and conclusions are summarized in the last section.

\section{Maxwell Demon-assisted EPR steering}
We will now consider how the Maxwell demon can be involved in the EPR steering task. In the standard EPR steering test \cite{steering1}, Alice sends qubits to Bob, who does not trust Alice and is unsure whether the qubits he has received are one-half of entangled pairs. After receiving the qubit, Bob will announce his choice of measurement setting and ask Alice to declare her result. Based on his measurement results and Alice's declared results over multiple runs, Bob can calculate the steering parameter $S_{m}$, where $m$ is the number of measurement settings, and check if $S_{m}$ is greater than a certain classical bound. If this is true, then Bob is convinced that he has received qubits from entangled pairs, and Alice has demonstrated steering of his state \cite{steering2}.
The demonstration of EPR steering refutes the local hidden state model that Alice could use to cheat in tests. However, there may be other cheating strategies.

Suppose that Alice chooses to collaborate with a Maxwell demon, and they make an agreement on the cheating strategy as follows. At the very beginning of the test, the Maxwell demon will accompany the qubit carrier, for example, photons entering Bob's device without being detected. The demon then secretly gains access to a specific part of Bob's device and tries to obtain information about the measurement setting. There are two situations in which Bob settles down his choice of measurement settings. The first one is to settle down many run choices before Alice sends the qubit and these classical data are stored in the local memory. The second one is to begin to choose one of the measurement settings after receiving the qubit sent by Alice and making the choice for every run. In the first case, the demon can directly memorize these classical data. For example in the case of two measurement settings, Bob's quantum randomness generator would generate classical bitstring like $0110\cdots$, which corresponds to different measurement settings chosen in each experimental run. These classical data have to be stored in the local memory of Bob's device. Since the classical bit can be cloned without changing its state, the demon can access the local memory and copy the bit information before Bob performs the qubit measurement.
For the second case, the demon could initially entangle with Bob's quantum randomness generator, causing the composite state to becomes
\begin{equation}
|D=0\rangle\otimes \frac{1}{\sqrt{m}}\sum_{k=1}^{m}|k\rangle\rightarrow\frac{1}{\sqrt{m}}\sum_{k=1}^{m}|D=k\rangle\otimes|k\rangle,
\end{equation}
where $|D=0\rangle$ represents the standard state of the demon and $m$ is the number of measurement settings. The orthogonal states $\lbrace |k\rangle\rbrace$ correspond to different measurement settings. Once Bob confirms his state $|k\rangle$ and thus his choice via measurement, the state of the demon collapses into the corresponding state $|D=k\rangle$. In both cases, the demon can obtain the choice information of Bob before he measures the qubit received. In the following discussion we only focus on the second case since it is what happens in practical experiments.
Once the demon knows Bob's choice, it will immediately perform a corresponding operation on the qubit before it is measured by Bob. The qubits sent by Alice are all prepared in the same pure state and will be transformed into eigenstates of Bob's chosen measurement setting by the demon. After the operation of the qubit, the demon informs Alice of its specific operation when a classical information channel is established between Alice and Bob. At last, Alice declares her result to Bob, based on the demon's information. The above process, as shown in Fig. 1, involves multiple runs, completing the Maxwell Demon-assisted EPR steering.

\section{Physical limitation of Maxwell demon}
The Maxwell Demon-assisted EPR steering described above suggests that quantum non-locality demonstrated by EPR steering can be simulated using local operations and classical communication. This Maxwell Demon loophole is a new type of loophole that goes beyond the existing ones, such as the detection loophole, locality loophole, and free will loophole. To possibly close this Maxwell Demon loophole, we need to consider the physical limitation of Maxwell Demon and utilize Landauer's erasure principle. 

\begin{figure}[tbp]
\centering
\includegraphics[scale=0.29]{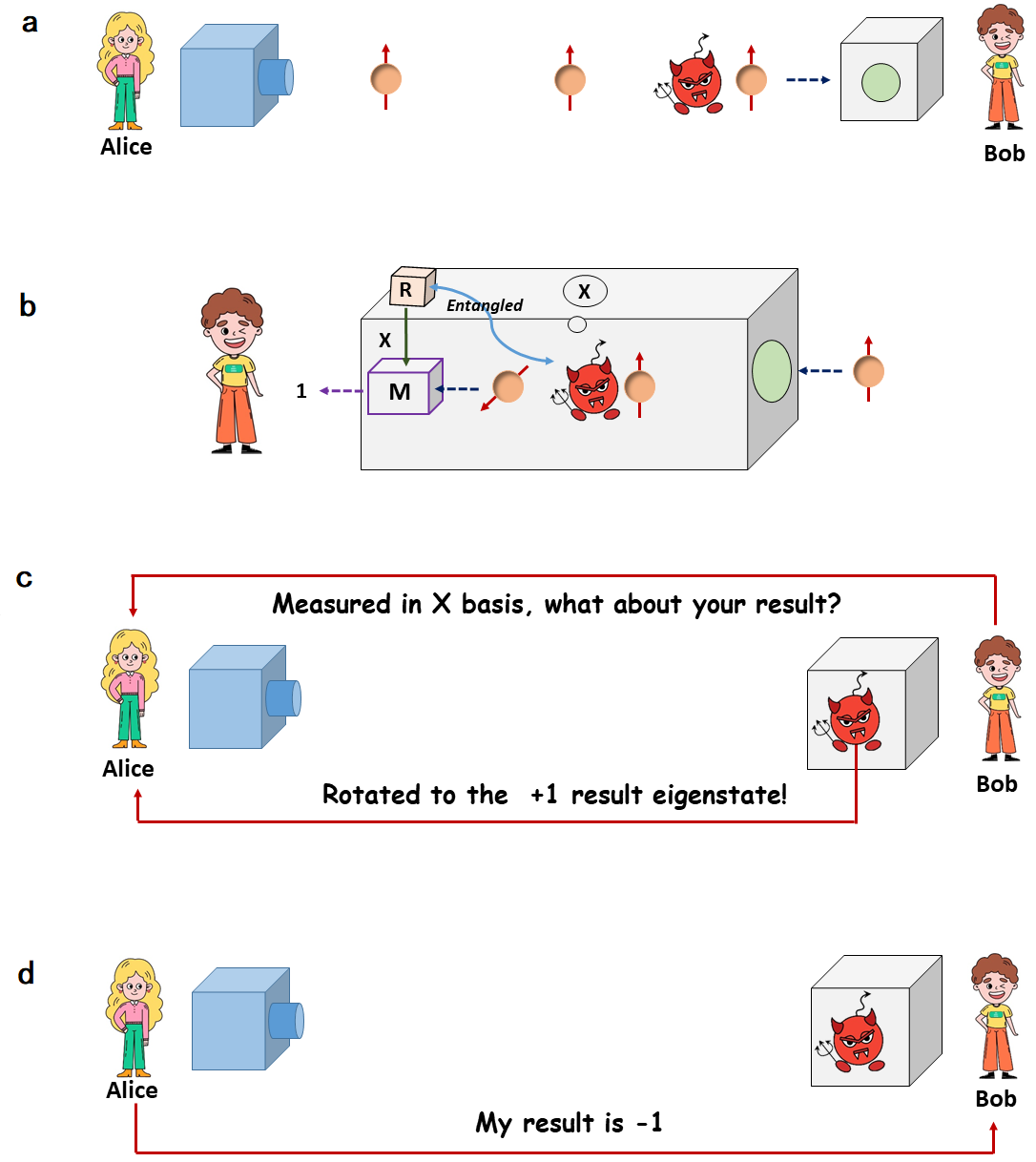}
\caption{Schematic diagram of Maxwell Demon-assisted EPR steering task. {\bf a:} Demon with qubits all prepared in the same state, e.g., $|0\rangle$ are sent to Bob by Alice. {\bf b:} The demon entangles itself with the local randomness generator of the Bob so that it can get access to the measurement basis choice of Bob. Before Bob performs measurement, the demon rotates the qubit to be measured into one of the eigenstates of measurement basis randomly. {\bf c:} When Bob tells Alice his measurement basis and asks for her result, the demon also informs Alice secretly which eigenstate it has rotated into for this measurement basis. This additional $1$ bit information is hidden in the classical communication between Bob and Alice, and it is reasonable to assume that it can not be detected by Bob. {\bf d:} Based on the information given by Bob and the demon, Alice answers back to Bob about her result on the same basis. {\bf a-d} repeat many times to complete the EPR steering task. }
\label{f1}
\end{figure} 

The key point is that the memory capacity of a physical Maxwell demon should be very limited. One might envision an artificial intelligence machine with a sufficiently large memory as a demon. However, this machine must be physically large enough compared to the qubit carrier, which can be easily detected by Bob. The physical size limitation of the demon implies that it has only limited memory capacity. After each run, as shown in Fig. 1, the demon has to erase its memory to recover its standard state $|D=0\rangle$ for the next runs. According to Landauer's erasure principle, there must be energy dissipated into the local environment. For the erasure of $1$ bit of information, the minimal energy to be dissipated is $k\mathrm{Tln2}$, with $k$ being the Boltzmann constant and $T$ being the temperature of the local environment. In the $N$ runs of the non-locality test, it is implied that at least $N\cdot k\mathrm{Tln2}$ of heat energy is dissipated in Bob's device in the EPR steering case. Under the room temperature condition $\mathrm{T=300}$ $\mathrm{K}$, at least $k\mathrm{Tln2}=2.87\times 10^{-21}$ $\mathrm{J}$ of energy will be dissipated into the local environment for the erasure of $1$ bit of information. The abnormal heat effect can, in principle, be detected by the local participant, i.e., Bob, given that the test runs with a sufficiently large value of $N$. The most advanced thermometer reported has a sensitivity that can record a change in heat energy of $10^{-6}$ $\mathrm{J}$ \cite{Therm}, which requires at least $N\approx 10^{15}$ test runs.
For the current state-of-the-art entangled photon pairs generator \cite{Pan}, the source rate can reach approximately $10^{9}$ $\mathrm{Hz}$. The experiment needs to run continuously for more than $12$ days to meet the minimum detectable requirement. 
To close this Maxwell demon loophole, Bob must be able to monitor its local environment at all times during the test. The energy fluctuations of the local environment must be controlled to a level that does not obstruct the detection of abnormal effects caused by the Maxwell demon.
This would be very demanding in practice, but it is not impossible with most advanced technologies.

\section{Quantum Circuit Model}
The Maxwell demon-assisted EPR steering can be demonstrated by programmable quantum processors, such as superconducting quantum computers \cite{Sup1, Sup2, Sup3, Sup4}. Here, we discuss in detail the quantum circuit realization as depicted in Fig. 2. Using this quantum circuit model, we establish the quantitative relationship between the work done by the demon and the quantum non-locality demonstrated in the EPR steering.  

\begin{figure}[tbp]
\centering
\includegraphics[scale=0.32]{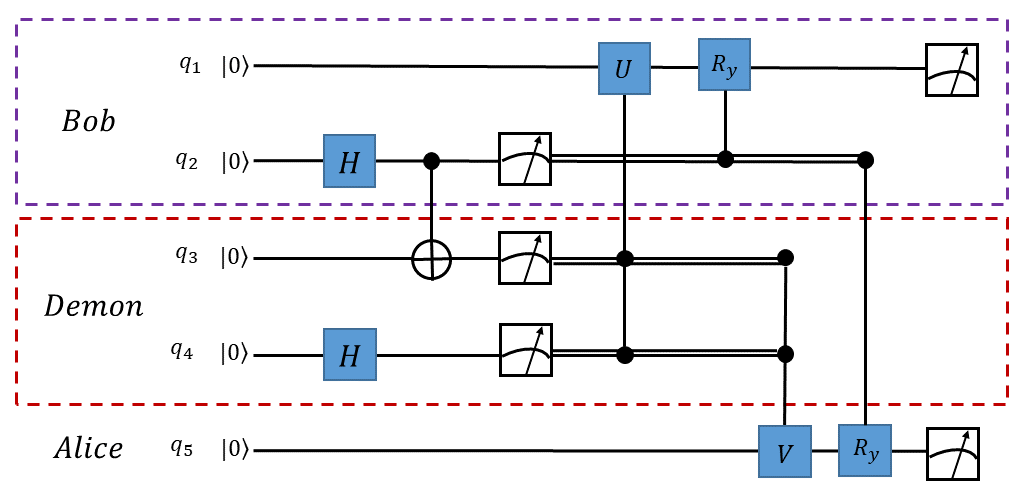}
\caption{Quantum circuit realization of Maxwell demon-assisted EPR-steering in the case of two measurement settings.  The demon consists of two qubits, in which $q_{3}$ entangles with the random generator of Bob $q_{2}$ by Control-Not gate to access the measurement basis information. The demon rotates the qubit $q_{1}$ randomly into one of the eigenstates of measurement basis via three-qubit classically Control-Control-U gate before Bob performs basis measurement. The specific function of this three-qubit gate is determined by Eq. (2). The response of Alice to Bob based on the demon's information is realized via another three-qubit classically Control-Control-V gate. For the simulation of positive correlation $\langle 0|V^{\dagger}U|0\rangle=1$, while $\langle 0|V^{\dagger}U|0\rangle=0$ for anti-correlation. It should be noted that the measurements on qubits $q_{2}, q_{3}, q_{4}$ can be moved to the end of the circuit without loss equivalence according to the principle of deferred measurement. In this case, the three-qubit classically controlled gate operation will be replaced by a three-qubit conditional quantum gate operation \cite{Nilson}.}
\label{f2}
\end{figure} 

The Maxwell demon, as shown in Fig. 2, consists of two qubits, namely $q_{3}$ and $q_{4}$, with $q_{4}$ serving as the random choice generator for the demon. Bob also has two qubits, with $q_{1}$ representing the qubit received from Alice, and $q_{2}$ serving as the random choice generator for his measurement basis. In our case, $q_{2}$ in the state of $|0\rangle$ or $|1\rangle$ corresponds to Pauli measurement basis $Z$ or $X$ respectively. At the beginning of process, $q_{3}$ of the demon becomes entangled with $q_{2}$ via a Control-NOT gate, resulting in their state becoming $(|0\rangle_{2}|0\rangle_{3}+|1\rangle_{2}|1\rangle_{3})/\sqrt{2}$. When Bob performs a measurement on $q_{2}$, $q_{3}$ will collapse into the same state as $q_{2}$, which implies the demon immediately knows the basis information of Bob. According to the measurement result of $q_{2}$ and $q_{4}$, the demon will perform the corresponding operation on $q_{1}$ before Bob measures the qubit with his selected basis. The purpose of this operation of the demon is to rotate the state of $q_{1}$ into one of the eigenstates of the measurement basis. The demon's operation can be realized by using a three-qubit classically Control-Control-$U$ gate, as shown in Fig. 2.
The specific rotation form of $U$ is determined by
\begin{equation}
U|a\rangle_{3}|b\rangle_{4}|0\rangle_{1}=|a\rangle_{3}|b\rangle_{4}\otimes e^{-i\frac{\pi}{4}(a+2b)\hat{\sigma}_{Y}}|0\rangle_{1},
\end{equation}
where $a, b\in \lbrace 0, 1\rbrace$ and $\hat{\sigma}_{Y}$ is a Pauli operator. \Com{}
As an example, $a=1$ implies that the measurement basis is chosen to be $X$, and the qubit $q_{1}$ be rotated into state $|+\rangle$ or $|-\rangle$ if $b=0$ or $b=1$, respectively. Here $|\pm\rangle=(|0\rangle\pm|1\rangle)/\sqrt{2}$ are eigenstates of $X$, i.e., $X|\pm\rangle=\pm|\pm\rangle$. After Bob performs the measurement and asks Alice for her result on the same basis, Alice responds based on the information sent by the demon. In the quantum circuit, this process is completed by a three-qubit classically Control-Control-V gate and Control-$R_{y}$. The specific form of $V$ depends on the type of correlation we want to simulate for an entangled pair in the EPR-steering. For a positive correlation, $V$ and $U$ should satisfy the relation $\langle 0|V^{\dagger}U|0\rangle=1$, while for anti-correlation, $\langle 0|V^{\dagger}U|0\rangle=0$. At last, Control-$R_{y}$ realizes different basis transformations.
It should be noted that we only consider the case of two measurement settings here, i.e., orthogonal Pauli measurement basis $Z$ and $X$. For more measurement settings, more qubits are required for the random generator. More additional qubits mean more multi-qubit gates need to be executed in the quantum circuit, which is a challenge for current quantum processors with limited circuit depth \cite{NISQ}. Since the underlying physics does not change, we will confine the discussion to the two measurement settings.

Given the quantum circuit model of Maxwell demon-assisted EPR-steering above, we are now ready to discuss the relationship between the work done by the demon and the quantum non-locality demonstrated in EPR-steering. The minimum demon, as shown in Fig. 2, consists of two qubits, namely $q_{3}$ and $q_{4}$. They have to be reset to the ground state $|0\rangle$ after each run. There are four possible states of $q_{3}$ and $ q_{4}$ with equal probability at the end of the quantum circuit, i.e., $|0\rangle_{3}|0\rangle_{4}$, $|0\rangle_{3}|1\rangle_{4}$, $|1\rangle_{3}|0\rangle_{4}$ and $|1\rangle_{3}|1\rangle_{4}$. For a qubit with the state $|1\rangle$, resetting the state to the ground state $|0\rangle$ is equivalent to erasing $1$ bit of information. The reset of $q_{3}$ and $ q_{4}$, on average, results in the erasure of $1$ bit of information for each run. According to Landauer's erasure principle, at least $k\mathrm{Tln2}$ energy is dissipated into the local environment due to the work done by the demon to erase $1$ bit of information in each run. In a quantum circuit, the lower bound of dissipated energy is determined by the difference between two energy levels of the physical qubit. Taking a superconducting qubit as an example, the natural way to reset the qubit is through spontaneous emission from the excited state $|1\rangle$ to the ground state $|0\rangle$, during which a microwave photon with energy $\hbar w_{q}=E_{1}-E_{0}$ is emitted to the local environment. The frequency $w_{q}$ of superconducting qubit is approximately 5 $\mathrm{GHz}$ \cite{Sup1}.
A superconducting quantum chip typically operates in a low-temperature environment with $T\approx$ 20 $\mathrm{mK}$ \cite{Sup2}, resulting in $k\mathrm{Tln2}\approx 10^{-25}\mathrm{J}\le \hbar w_{q}$. 
In practice, however, additional work must be done to speed up the reset rate, resulting in more energy being dissipated into the environment. Since the operation is the same during the reset process, we can reasonably assume that the dissipated energy due to the work of the demon $E_{D}\ge\hbar w_{q}$ is the same for each run. If the demon does operations on Bob's qubit $q_{1}$ for each run to simulate an entangled pair $|\Psi\rangle$, the total energy dissipated for $N$ runs will be $N\cdot E_{D}$. In a more realistic scenario, the demon could perform operations with a probability $p$ for each run. This means that the simulated entangled pair of qubits $q_{1}$ and $q_{5}$ for $N$ runs is described by 
\begin{equation}
\rho_{15} = p|\Psi\rangle_{15}\langle\Psi| + (1-p)|0\rangle_{1}\langle 0|\otimes|0\rangle_{5}\langle 0|,
\end{equation}
where the specific form of $|\Psi\rangle_{15}$ depends on the type of correlation being simulated, for example, $|\Psi\rangle_{15}=(|0\rangle|1\rangle-|1\rangle|0\rangle)/\sqrt{2}$ for the case of anti-correlation. 
The steering parameter in the case of two measurement settings is given by
\begin{equation}
\begin{split}
S_{2}(\rho_{15})&=\frac{1}{2}\lbrace|\mathrm{Tr}[(\hat{\sigma}_{z}\otimes\hat{\sigma}_{z})\rho_{15}]|+|\mathrm{Tr}[(\hat{\sigma}_{x}\otimes\hat{\sigma}_{x})\rho_{15}]|\rbrace \\
&=\frac{1}{2}(1+p),
\end{split}
\end{equation}
which is independent of the type of correlation being simulated. The successful demonstration of EPR steering requires that the obtained steering parameter $S_{2}$ should be larger than the corresponding classical bound $C_{2}$. If $\Delta_{2}\equiv S_{2}-C_{2}> 0$ is defined as the measure of quantum non-locality demonstrated in EPR steering, we obtain the following equation
\begin{equation}
E(\Delta_{2})=pNE_{D}=[2(\Delta_{2}+C_{2})-1]NE_{D}
\end{equation}
with $C_{2}=1/\sqrt{2}$ and $E_{D}\ge\hbar w_{q}$. The equation quantitatively describes the relationship between the work done by the demon and the quantum non-locality demonstrated in EPR steering.  In the extreme case that $m\gg 1$, $S_{m}(\rho_{15})\to p$ and $C_{m}\to 1/2$, and we have $E(\Delta_{m})\to (\Delta_{m}+1/2)NE^{m}_{D}$. The linear relation given in Eq. (5) holds for the case of measurement settings $m>2$, where $E(\Delta_{m})\varpropto\Delta_{m}NE^{m}_{D}$.

\section{Discussion and Conclusion}


Quantum non-locality demonstrated in EPR steering is strictly weaker than that in the Bell test. We may wonder if the above discussion of Maxwell demon applies to the case of the Bell test. The answer, unfortunately, is negative. To make it explicit, let's consider a Bell game in which a third party, Charlie, who chooses collaboration with the demons, tries to simultaneously deceive Alice and Bob, who are separated by distant space, into believing that their received qubits are entangled. In order to achieve this goal, two demons have to be sent secretly to Alice and Bob, respectively. The operations of demons are almost the same in the case of EPR steering, i.e., they have access to measurement basis information and rotate the qubits before they are measured. The main issue, however, is how to rotate the qubits based on the basis information. In EPR steering, the demon can randomly rotate the qubit into one eigenstate of measurement basis due to the classical communication between Alice and Bob, and the demon could inform Alice of its operation. This is not possible in the Bell test, in which classical communication is prohibited. One might imagine that two demons could share a table list from the beginning, guiding them on how to operate based on the obtained basis information. That would require a table list long enough to cover all the runs, which is physically impossible. One may also imagine that the random generators of two demons are entangled, such that their operations are correlated. This can indeed be possible using the quantum circuit model. However, it is a logical circular argument because it uses the non-locality correlation of demons to replace the non-locality correlation between Alice and Bob in the Bell test. We thus conclude that there is no Maxwell demon loophole in the Bell test.

In conclusion, we have addressed the issue of simulating quantum non-locality through work. In the task of EPR steering, the Maxwell demon can be introduced in collaboration with Alice to deceive Bob using only local operations and classical communication. The existence of Maxwell demon-assisted EPR steering implies a new-type loophole, i.e., {\it Maxwell demon loophole}, which can only be closed by carefully monitoring heat fluctuations in the local environment by the participant. To give a quantitative relationship between quantum non-locality correlation in EPR steering and the work done by the demon, we construct a quantum circuit model of Maxwell demon-assisted EPR steering, which can be demonstrated in current quantum processors.
We show that there exists a simple relation between the energy dissipated in the local environment due to the work of the demon and the violation of classical local correlation. 
Our results provide a new approach to exploring and better understanding of relationships between quantum non-locality, information theory, and thermodynamics.

{\it Note added}: Experimental demonstration of Maxwell demon-assisted EPR steering via superconducting quantum processor has been reported recently in \cite{Ex}.

\hfill

\begin{acknowledgments}
{\bf Acknowledgments}: Meng-Jun Hu extends special thanks to Peng Zhao for highlighting the potential use of superconducting quantum circuits in studying the Maxwell demon model. The authors acknowledge Paul Skrzypczyk for valuable suggestions.
Meng-Jun Hu acknowledges the support from the Natural Science Foundation of China (Grant No. 92365206). Xiao-Min Hu acknowledges the support from the Fundamental Research Funds for the Central Universities, and Yong-Sheng Zhang is supported by the National Natural Science Foundation of China (Grant No. 92065113).

\end{acknowledgments}

\end{document}